\documentstyle[11pt,aaspp4]{article}

\def\lsim{\lower.5ex\hbox{$\; \buildrel < \over \sim \;$}}
\def\gsim{\lower.5ex\hbox{$\; \buildrel > \over \sim \;$}} 
\def\lax    {\ifmmode{_<\atop^{\sim}}\else{${_<\atop^{\sim}}$}\fi}
\def\gax    {\ifmmode{_>\atop^{\sim}}\else{${_>\atop^{\sim}}$}\fi}

\def\gtorder{\mathrel{\raise.3ex\hbox{$>$}\mkern-14mu
	     \lower0.6ex\hbox{$\sim$}}}
\def\ltorder{\mathrel{\raise.3ex\hbox{$<$}\mkern-14mu
	     \lower0.6ex\hbox{$\sim$}}}

\begin{document}

\title{Monte Carlo Simulation of Comptonization in Inhomogeneous Media}

\author{Xin-Min Hua\altaffilmark{1}} 
\affil{Laboratory for High Energy Astrophysics, NASA/GSFC Code 661, 
       Greenbelt, MD 20771}
\affil{hua@rosserv.gsfc.nasa.gov}
\altaffiltext{1}{also Universities Space Research Association}

\vskip 2.0 truecm

\centerline{ABSTRACT}

Comptonization is the process in which photon spectrum changes due to
multiple Compton scatterings in the electronic plasma. It plays an
important role in the spectral formation of astrophysical X-ray
and gamma-ray sources. There are several intrinsic
limitations for the analytical method in dealing with the 
Comptonization problem and Monte Carlo simulation is one 
of the few alternatives. We describe an efficient Monte Carlo method 
that can solve the Comptonization problem in a fully relativistic way. 
We expanded the method so that it is capable of simulating 
Comptonization in the media where electron density and temperature
varies discontinuously from one region to the other and in the isothermal
media where density varies continuously along photon paths. The algorithms 
are presented in detail to facilitate computer code implementation. We 
also present a few examples of its application to the astrophysical research. 

\section{Introduction} 

Comptonization -- the process where photon spectrum changes due to
multiple Compton scatterings in the electronic plasma -- is one of
the most important processes in the spectral generation of 
X-ray binaries, active galactic nuclei and other X-ray and 
gamma-ray sources. The analytical treatment of Comptonization 
are essentially based on the solution of Kompaneets equation which
describes the interactions between radiation field and thermal 
electrons (Kompaneets 1956). Due to the mathematical complexity,
however, previous analysis of Comptonization depended on simplifications
such as the non-relativistic approximation and therefore the results 
were only applicable to a relatively small range of photon and electron 
energies (e.g. Sunyaev \& Titarchuk 1980). In recent years, 
Titarchuk (1994) developed a modified analytical technique which 
took into account the relativistic effect and Klein-Nishina 
corrections, thereby extending the previous work to wider ranges of 
temperature and optical depth of the plasma clouds from which Comptonized 
photons emerge. 

The analytical method, however, have several intrinsic limitations. 
First, all analytical models are based on 
solving certain types of radiation transfer equations (Kompaneets 1956),
which in turn is based on the assumption that energy and position of the 
photons are continuous functions of time, i.e. these models assume
diffusion of photons in the energy and position spaces. 
While the continuity of energy 
change is a good approximation for scatterings at low energy, it is 
obviously not valid for Compton scatterings at high photon energies or 
by relativistic electrons. Similarly, the continuity of photon position 
change is an approximation only valid for clouds of electron plasma
with dimensions large compared to the scattering mean free path (i.e.
diffusion approximation). But astronomical
observations suggest that many of the sources where Comptonization is
believed to take place have optical depths of the order of one 
Thomson scattering mean free path.     

Second, solutions of the radiative transfer equations are based on the
separation of photon diffusions in energy and position spaces
(Sunyaev \& Titarchuk 1980, Titarchuk 1994 and Hua \& Titarchuk 1995).
The solutions can be presented in terms of simple analytical
expressions only when initial source photons have energies much lower
than the electron energy and follow a particular spatial distribution, namely, 
the first eigenfunction of the spatial operator of the diffusion 
equation. It was found (Hua \& Titarchuk 1995) that for source photons 
at energies not far below the electron energy or for clouds with 
large optical depth, the emergent spectra are sensitive to both the 
spectral and spatial distributions of source photons and the results 
of analytical method must be expanded to the higher order terms. 
Consequently, the analytical models are 
applicable only to certain ranges of plasma temperature and optical 
depth where solutions are insensitive to source conditions.

Third, the analytical methods are inadequate to treat the temporal
behavior of Comptonized emissions. Hua \& Titarchuk (1996) have shown
that for relativistic plasma, photons gain energy significantly with each 
scattering and consequently the scattering mean free path changes
significantly with each scattering. Besides, for plasma clouds with 
small optical depth, the scattering mean free path are mainly 
determined by the boundary condition instead of the scattering cross sections. 
As a result, analytical treatment (e.g. Payne 1980), is only 
applicable to the limited situation in which electron plasma 
has non-relativistic temperatures and optical depths much 
greater than Thomson mean free path. 

In addition to the above limitations, analytical approach is totally
incapable of dealing with the Comptonization problems involving complicated 
geometries and inhomogeneity of electronic media, where scattering mean
free path depends on scattering location and direction as well as photon 
energy. But observations seem to indicate that investigations of 
Comptonization in the 
media with non-uniform temperature and density are necessary.  
As was shown by Skibo et al. (1995) and Ling et al. (1997), the
spectral hardening at high energies in the spectra of AGNs and black
hole candidates may be resulting from the temperature gradient in
the plasmas responsible for the emissions. Kazanas et al. (1997) and Hua et
al. (1997) showed that the temporal behavior such as the hard X-ray
phase lags observed from the accreting compact objects may be explained 
by the non-uniform electron density of the accreting gas clouds.

These situations are where analytical method fails. As an alternative,
Monte Carlo simulation can be employed to give solutions. It is flexible in 
simulating various initial conditions of source photons, complicated 
geometries and density profiles of plasma clouds. It is capable of 
presenting the full spectra resulting from Comptonization rather than 
the asymptotic ones obtainable from analytical methods. The first
attempt to use Monte Carlo method to solve Comptonization
problem was by Pozdnyakov et al. (1983). In recent years, Stern et al.
(1995) presented a large-particle Monte Carlo method for simulating 
Comptonization and other high-energy processes. Skibo et al. (1995)
used a Monte Carlo simulation in the calculation of photon spectra of
mildly relativistic thermal plasmas in pair balance.  

In this study, we develop an efficient Monte Carlo method which treats 
Comptonization problem in a fully relativistic way and can be 
implemented in a medium computer such as Sparc workstation or Pentium 
PC to yield results with satisfactory statistics in CPU time of the 
order of minutes to hours. The algorithms are described in detail to
facilitate computer code implementation.
In \S 2 we introduce an improved technique of simulating 
Compton scattering of photons on cold electrons. In \S 3, we describe
the method for Compton scattering on hot electrons. In \S 4, we present 
the method dealing with scattering in multi-zone medium. In \S 5,
we describe the simulation of Compton scatterings in media with non-uniform
density profiles.
 
\section{Compton Scattering on Cold Electrons}

The Monte Carlo method described here was developed over the past 
several years in the investigations of Compton scattering of 2.223 MeV 
gamma-ray line in solar flares (Hua 1986), Compton backscattering of 
511 keV annihilation line in the sources 1E1740.7-2942 (Lingenfelter 
\& Hua 1991) and Nova Muscae (Hua \& Lingenfelter 1993). 

The differential cross section of Compton scattering is given by
the Klein-Nishina formula  
$${d\sigma \over d\varepsilon} = {3\sigma_T \over 4} \cdot 
{1 \over \varepsilon} \left[\left(1-{4 \over \varepsilon}-
{8 \over \varepsilon^2}\right)\ln (1+\varepsilon) +{1 \over 2} +
{8 \over \varepsilon} -
{1 \over 2(1+\varepsilon)^2}\right], \eqno(1)$$

\noindent
where $\sigma_T$ is Thomson cross section; $\varepsilon = 2E/m_ec^2$; 
$E$ is the energy of incident photon; $m_e$ is the electron rest
mass and $c$ the speed of light. The energy of the scattered photon, 
$E'$, relative to the initial photon energy $E$ is given by the ratio
$$r = \displaystyle{E \over E'} = 1+\displaystyle{\varepsilon \over 2}
(1 - \cos\psi), \eqno(2)$$

\noindent
where $\psi$ is the angle between incident and scattered photons.
The energy distribution of the Compton-scattered photons is determined
by the distribution with respect to $r$, which is
$$f(r) = \cases {\displaystyle{{1}\over {K(\varepsilon)}}\left[\left(
\displaystyle{{\varepsilon +2 -2r}\over {\varepsilon r}}\right)^2+
\displaystyle{{1}\over {r}} - \displaystyle{{1}\over {r^2}}+ 
\displaystyle{{1}\over {r^3}} \right]  
&{\rm for $1\leq r \leq \varepsilon+1$}, 
\cr 0  &{\rm otherwise},} \eqno(3)$$
\noindent
where 
$$K(\varepsilon) = \frac{4\varepsilon}{3\sigma_T} \sigma(\varepsilon) 
\eqno(4)$$

\noindent
is the normalization factor.

Sampling the distribution given by Eq. (3) plays a central role 
in the Monte Carlo simulation of Compton scattering of photons by cold 
electrons. Furthermore, as will be seen below, Compton scatterings on 
hot electrons in our scheme will also be reduced to the simulation of
Eq. (3). Therefore, the performance of the computer program for Monte 
Carlo simulation of Compton scatterings depends critically on the 
quality of the technique used for sampling this distribution because 
a run of the program typically involves millions of scatterings. 
Efforts were made to optimize the technique of sampling this
distribution (e.g. Kahn, 1954). In our implementation, we adopted a
variation of Kahn's technique first suggested by Pei (1979). The
algorithm of the technique is
\begin{tabbing}
1. \= Generate 3 random numbers $\xi_1, \xi_2$ and $\xi_3$ uniformly
distributed on (0,1). \\
2. \= If \= $\xi_1 \leq 27/(2\varepsilon + 29)$, \\
   \>    \> let $r = (\varepsilon +1)/(\varepsilon \xi_2 +1)$. \\
   \>    \> If $\xi_3 > 
            \{[(\varepsilon +2 -2r)/\varepsilon]^2 + 1\}/2$, go to 1. \\
   \>    \> Else accept $r$. \\
   \> Else \\
   \>    \> let $r = \varepsilon \xi_2 +1$. \\
   \>    \> If $\xi_3 > 6.75 (r -1)^2/r^3$, go to 1. \\
   \>    \> Else accept $r$.
\end{tabbing}
It is seen that this is essentially a combination of composition and
rejection methods (see e.g. Ripley, 1987). 
This algorithm, like Kahn's, avoids the operations such as
square root, logarithm or trigonometric functions, which involve
time-consuming series expansion for computers.  
Its quality can also be measured to a large extent
by the rejection rates, which are 0.38, 0.30, 0.23 and 0.33 for 
$\varepsilon = 0.2, 2, 10$ and 20 respectively,
as compared to 0.41, 0.37, 0.41 and 0.53 for Kahn's technique. The 
improvement is significant, especially for higher photon energies. 

\section {Comptonization in Hot Isothermal Homogeneous Plasmas}

The Monte Carlo technique for photon Comptonization in a relativistic
plasma was outlined by Pozdnyakov et al. (1983) and Gorecki \&
Wilczewski (1984). Our implementation of the simulation is somewhat
different from these authors. It was developed on the bases of
the technique for Compton scattering on cold electrons described in 
the last section.  

Suppose a photon is scattered off an electron which is moving
in $z$-axis direction with a velocity $v$. The energies of the
incident and the scattered photon are $E$ and $E'$ respectively.
The zenith angles of the incident and scattered photons measured
from $z$-axis are $\theta~ {\rm and}~ \theta'$ respectively. $\phi$ and
$\phi'$ are the azimuthal angles. The differential
cross section for Compton scattering is given by (see e.g. Akhiezer \&
Berestetskii 1969)
$${d\sigma \over d\mu'd\phi'} = {3\sigma_T \over 16\pi} {1 \over
\gamma^2}
{\chi\over (1-v\mu)^2} \left({E' \over E}\right)^2,
\eqno(5) $$

\noindent
where $\mu=\cos\theta$ and $\mu'=\cos\theta'$;
$v$ is in units of the speed of light and $\gamma = (1-v^2)^{-1/2}$;
$$\chi={\varepsilon \over \varepsilon'}+
{\varepsilon' \over \varepsilon}+{4 \over \varepsilon}\left(1-
{\varepsilon \over \varepsilon'}\right)+
{4 \over \varepsilon^2}\left(1-{\varepsilon \over
\varepsilon'}\right)^2; \eqno(6)$$
$$\varepsilon={2E \over m_e c^2}\gamma(1-v\mu), \qquad
\varepsilon'={2E' \over m_e c^2}\gamma(1-v \mu'); \eqno(7)$$
$${E' \over E} = {{1-v\mu} \over {1-v\mu'+
(E/\gamma m_ec^2)(1-\cos\psi)}}; \eqno(8)$$

\noindent
and $\psi$ is the angle between incident and scattered photons
$\cos\psi = \mu\mu'+\sqrt{(1-\mu^2)(1-\mu'^2)}\cos(\phi -\phi')$. 

Integration over $\mu'$ and $\phi'$ leads to
$$\sigma(\varepsilon)= {3\sigma_T \over 4} \cdot {1 \over \varepsilon}
\left[\left(1-{4 \over \varepsilon}-
{8 \over \varepsilon^2}\right)\ln (1+\varepsilon) +{1 \over 2} +
{8 \over \varepsilon} -
{1 \over 2(1+\varepsilon)^2}\right]. \eqno(9)$$

\noindent
It is seen that Eq. (9) is identical in form with Eq. (1).
But the quantity $\varepsilon$ here is given by the relativistic
expression in Eq. (7). In other words, it is dependent on the
electron's energy and direction as well as  photon's energy. 

A photon with energy $E$  traveling in a plasma with an isotropic 
distribution of electrons having an energy distribution $N_e(\gamma)$ 
will have an averaged cross section of Compton scattering (see e.g. Landau \&
Lifshits, 1976):
$$\sigma_a(T_e,E)={1 \over 2}\int_1^{\infty}d\gamma\int_{-1}^1d\mu 
(1-v \mu)\sigma(\varepsilon) N_e(\gamma). \eqno(10)$$

\noindent
For a plasma in thermal equilibrium, $N_e(\gamma)$ is the
Maxwell distribution given by
$$N_e(\gamma) = {1 \over 2 \Theta K_2(1/\Theta)} v 
\gamma^2 e^{-\gamma/\Theta}, \eqno(11)$$

\noindent
where $\Theta = kT_e/m_ec^2$ is the dimensionless temperature of the
plasma; $k$ is the Boltzmann constant and $K_2$ is the modified Bessel 
function of 2nd order. The $\sigma_a(T_e,E)$ values in the form of 
a data matrix, obtained by the 2-dimensional integration in Eq. (10) 
for a properly spaced array of $T_e$ and $E$, can be read by or
incorporated into the computer codes.  Values of $\sigma_a(T_e,E)$ for
several temperatures are numerically calculated and plotted in Figure 1. 
The dashed curve in the figure is the cross section at $T_e=0$, given
by the Klein-Nishina formula in Eq. (1). It can be seen that
for energetic photons scattering off the high temperature electrons,
the cross section can be smaller by a factor of 2 or more than off  
the cold electrons. In other words, hot plasmas are more transparent
than cold ones for photons. This has important effect on the
energy spectra emerging from such plasmas, which Titarchuk (1994) took
into account in his modification of the previous analytical results.
Its effect on the temporal behavior of X-ray and gamma-ray emission from
these plasma is even more significant and was discussed in Hua \& Titarchuk
(1996). 

\begin{figure}[ht]
\plotone{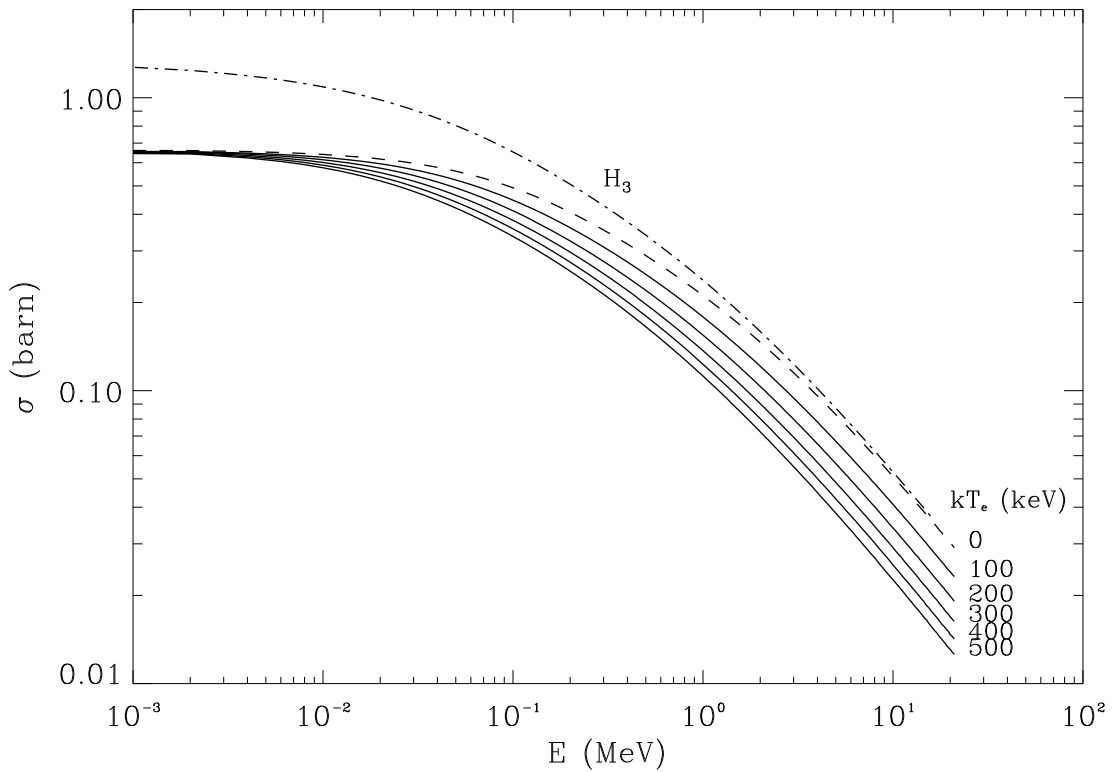}
\caption{
Maxwellian averaged Compton scattering cross section for
various plasma temperatures, obtained from numerical integration in
Equation (10). Also plotted is the maximum effective
cross section as a function of photon energy $H_3(E)$.}
\end{figure}

With $\sigma_a(T_e,E)$ obtained by numerical integration in Eq. (10), 
we can use the Monte Carlo method to select the free
path between two successive scatterings for a photon with energy $E$.
$$ \int_0^{\ell} n_e\sigma_a ds = -\ln \xi, \eqno(12)$$

\noindent
where $\ell$ is the free path to be sampled; $n_e$ is the electron
density and $\xi$ is a uniform random number on (0,1). The integration
is taken along photon's path length $s$. In this section
we are only concerned with the isothermal plasmas at temperature $T_e$ 
and with uniform density $n_e$ and leave the discussion about 
inhomogeneous media to the next two sections. Under this assumption, 
$\ell$ can be sampled simply by 
$$\ell = -\displaystyle{{\ln \xi}\over{n_e\sigma_a(T_e,E)}}. \eqno(13)$$

At the location of scattering, an electron is selected to scatter the
photon. Its energy factor $\gamma$ and direction $\mu =\cos \theta$ 
with respect to the photon direction are selected according to the 
distribution
$$f_e(\gamma,\mu)\propto (1-\mu \sqrt{1-\gamma^{-2}})
\sigma(\varepsilon)N_e(\gamma), \eqno(14)$$
\noindent
while its azimuthal angle $\phi$ around the photon direction is selected
uniformly on ($0, 2\pi$). The distribution in Eq. (14) is rather 
complicated because $\varepsilon$ depends on $\gamma$ and $\mu$ as given 
in Eq. (7). On the other hand, for a thermal plasma, 
$N_e(\gamma)$ is given by 
Eq. (11) and independent of $\mu$. In our implementation of the
distribution Eq. (14), we use the following algorithm.
\begin{tabbing}
1. \= Generate 2 random numbers $\xi_1$ and $\xi_2$ uniformly
distributed on (0,1). \\
2. \= If \= $\Theta \leq 0.01$, \\
   \> \> If \= $\xi_2^2  > -e \xi_1 \ln{\xi_1} $, go to 1. \\
   \> \> Else let 
         $v = \sqrt{-3 \Theta \ln \xi_1}$, $\gamma = 1/\sqrt{1-v^2}$. \\
   \> Else  if $\Theta \leq 0.25$, \\
   \>    \> let $\gamma = 1-1.5\Theta \ln{\xi_1}$. \\
   \>    \> If $\xi_2 H_1 >
            \gamma \sqrt{\xi_1(\gamma^2 -1)}$, go to 1. \\
   \> Else \\
   \>    \> let $\gamma = 1-3\Theta \ln{\xi_1}$. \\
   \>    \> If $\xi_2 H_2 >
            \gamma \xi_1^2 \sqrt{\gamma^2 -1}$, go to 1. \\
3. \> Generate $\mu$ uniform on (-1,1) and $\xi_3$ uniform on (0,1). \\
4. \> Calculate $\varepsilon$ and then $\sigma(\varepsilon)$ from
$\gamma$ and $\mu$ according to Eqs. (7) and (9). \\
5. \> If $\xi_3 H_3 > (1-\mu \sqrt{1-\gamma^{-2}}) \sigma(\varepsilon)$,
   go to 1.\\
   \> Else accept $\gamma$ and $\mu$.
\end{tabbing}
\noindent
Here 
$$H_1 = a \sqrt{a^2-1}~ \exp\left({-\displaystyle{{a-1}\over {3\Theta}}}
\right),$$
$$a = 2\Theta + 2b \cos\left[\displaystyle{1\over 3} \cos^{-1}
      \left(\Theta \displaystyle{{16\Theta^2 -1} \over {2 b^3}}
      \right)\right] ~~{\rm and}~~  b = \sqrt{1/3 +4 \Theta^2}; \eqno(15)$$
$$H_2 = a \sqrt{a^2-1}~ \exp\left[{-\displaystyle{{2(a-1)}\over {3\Theta}}}
\right],$$
$$a = \Theta + 2b \cos\left[\displaystyle{1\over 3} \cos^{-1}
      \left(\Theta \displaystyle{{4\Theta^2 -1} \over {4 b^3}}
      \right)\right]~~ {\rm and}~~ b = \sqrt{1/3 + \Theta^2}; \eqno(16)$$

\noindent
and $H_3$ is the maximum of the so called ``effective cross section"
$\sigma_{\rm eff}=(1- \mu \sqrt{1-\gamma^{-2}}) \sigma(\varepsilon)$. 
    
Several points should be made in the above algorithm. 
Steps 1 and 2 sample $\gamma$ using rejection
method in terms of the Maxwellian distribution $N_e(\gamma)$, which is 
independent of the photon energy and direction. For low plasma 
temperature ($\Theta \leq 0.01$) electron velocity $v$ are sampled 
according to the non-relativistic Maxwellian distribution. For high 
temperatures, the separated sampling ($\Theta \leq 0.25$ and $> 0.25$) is 
in order to reduce the rejection rates. It should be emphasized that although
the expressions of $H_1$ and $H_2$ are complicated, these quantities
depend on $\Theta$ alone and therefore need to be calculated once only.
They can be calculated outside the scattering loop as long as plasma 
temperature remains unchanged. The $\gamma$ values so sampled, together
with the isotropically sampled $\mu$, represent electrons in the hot
plasma at the given temperature. They are subject to another rejection
test in the subsequent steps in order to yield the right joint 
distribution given in Eq. (14), which represents the electrons that
actually scatter the photon. 

temperature remains unchanged. The $\gamma$ values so obtained are 
subject to another rejection test in the subsequent steps together with 
the isotropically sampled $\mu$ in order to yield the right joint 
distribution given in Eq. (14). 

The quantity $H_3$ is not expressible analytically. It depends on
incident photon energy $E$ only and can be determined by maximizing 
the effective cross section with respect to $\gamma$ and $\mu$ for 
any given $E$ using numerical methods such as given in Press et al.
(1992). In the following, we describe an alternative to the above 
2-dimensional maximization methods. We examine the derivative of 
$\sigma_{\rm eff}(\gamma,\mu)$ with respect to $\mu$
$$ {\partial\sigma_{\rm eff}\over \partial\mu} = -~{2E \over m_ec^2}
\sqrt{1-\gamma^{-2}}~{dh \over d\varepsilon}, \eqno(17)$$
\noindent
where 
$$h(\varepsilon) = \left(1-{4 \over \varepsilon}-
{8 \over \varepsilon^2}\right)\ln (1+\varepsilon) +{1 \over 2} +
{8 \over \varepsilon} - {1 \over 2(1+\varepsilon)^2} \eqno(18)$$
\noindent
is the expression in the square parentheses in Eq. (9). It can be
easily verified that 
$$ {\partial\sigma_{\rm eff}\over \partial\mu} \leq 0 
\qquad {\rm for}~~E > 0~~{\rm and}~~\gamma \geq 1 \eqno(19)$$
\noindent
Therefore, $\sigma_{\rm eff}(\gamma,\mu)$ is a monotonously decreasing 
function of $\mu$, that is, for given $\gamma$, $\sigma_{\rm eff}
(\gamma,\mu)$ reaches its maximum at $\mu = -1$. Physically, this means 
that head-on collision between the photon and electron always
has the maximum probability. Thus, in order to determine the maximum of 
$\sigma_{\rm eff}$ as a function of $\gamma$ and $\mu$, one only 
needs to maximize the one dimensional function $\sigma_{\rm eff}
(\gamma,-1)$. The maximum of $\sigma_{\rm eff}$, or $H_3$, as a
function of $E$ determined in this way is plotted in Figure 1 as 
the dash-dotted curve. It is seen that for high photon energies, 
the maximum effective cross section approaches the Klein-Nishina cross
section while at low energies it approaches twice the Thomson cross 
section. The $H_3$ values for an array of properly spaced $E$ values 
can be tabulated and incorporated into the computer codes. 

With the selected electron energy and direction represented by $\gamma$,
$\mu$ and $\phi$ uniform on $(0,2\pi)$, we proceed to determine the 
energy and direction of the scattered photon. In order to do so, we 
simulate Compton scattering in the frame where the electron before 
scattering is at rest rather than sampling the multivariate distribution 
of $E'$, $\mu'$ and $\phi'$ from Eq. (5).
The Lorentz transformation of the photon momentum between this 
reference frame and the lab frame is given by
$${\bf p'}={\bf p}-p[\gamma v-(\gamma-1){\bf \hat p}
\cdot {\bf \hat v}] {\bf \hat v}, \eqno(20)$$
\noindent
where ${\bf p}$ and ${\bf p'}$ are photon momentum vectors before and after
the transformation; ${\bf \hat p}$ and ${\bf \hat v}$ are unit vectors of
the photon momentum and electron velocity respectively.

In the electron rest frame, we utilize the Monte Carlo method
described in \S 2. The resulting momentum of the scattered photon 
is then transformed back to the lab frame using the same Eq. (20) 
with a reversed ${\bf \hat v}$. The energy
and direction of the scattered photon obtained in this way automatically
satisfy the energy conservation relationship given in Eq. (8).

As a crucial test we ran the program in which low frequency
photons were allowed to scatter in an infinite plasma at a given
temperature for a sufficiently long time. It was expected that the 
photon energy should approach the Wien distribution at the given plasma 
temperatures. One example of such evolution, the photon energy 
distribution recorded at varies times in a plasma of $kT_e = 200$ keV are
shown in Figure 2. It does approach the Wien form.

\begin{figure}[ht]
\plotone{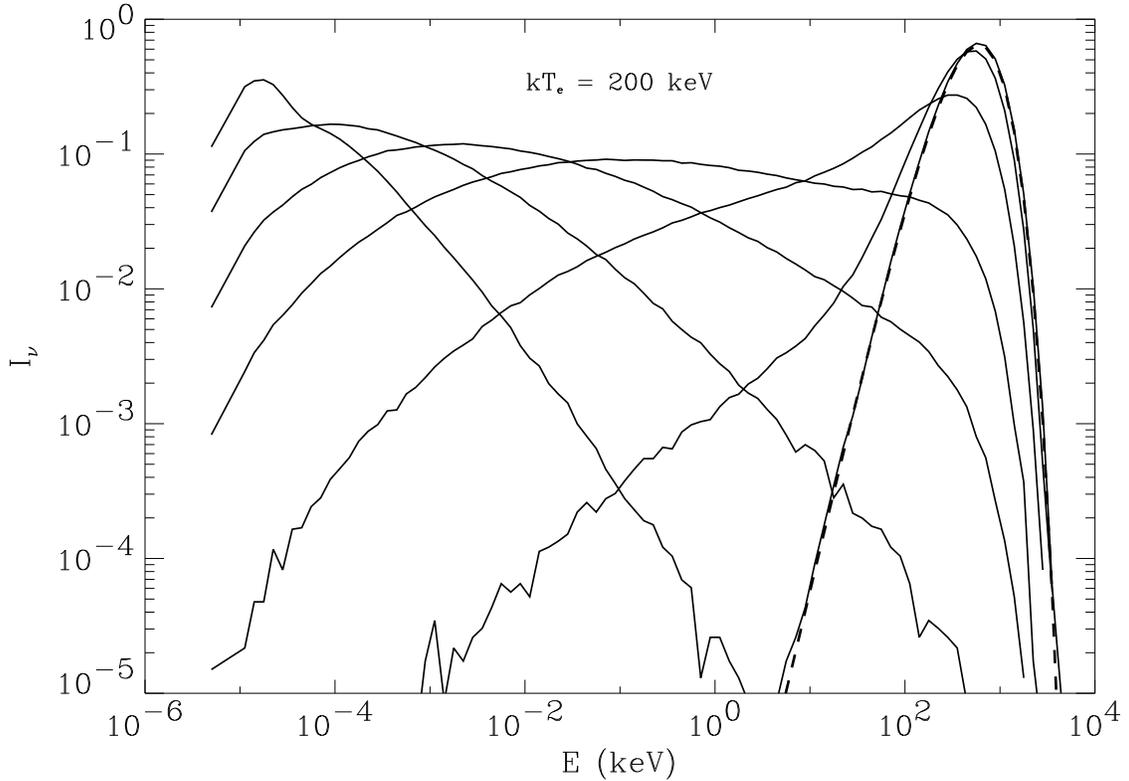}
\caption{
The evolution of photon energy spectrum from a blackbody
at $0.511 \times 10^{-2}$ eV towards equilibrium with a plasma at
$kT_e = 200$ keV. The seven spectra (solid curves) are ``snapshots''
at times $t = 1, 3, 6, 10, 18, 30, 70$ Thomson mean free time. Also
plotted (dashed curve) is the Wien spectrum at temperature 200 keV.}
\end{figure}

\section{Comptonization in Multi-Zone Media}

If Comptonization takes place in a medium which is divided into several
zones each with different electron temperatures and density distributions,  
one has to take into consideration the boundaries between these zones
in addition to the scattering free paths and the boundary of the entire 
medium. 

In general, suppose a photon, after initiation or scatterings, is located 
in the medium at $(x_0,y_0,z_0)$ with a direction $(\omega_1, \omega_2,
\omega_3)$. The next position where the photon will scatter, if there
were no boundaries, is given by
$$\cases {x_1=x_0+\ell \omega_1 \cr  
          y_1=y_0+\ell \omega_2 \cr 
          z_1=z_0+\ell \omega_3, } \eqno(21) $$  

\noindent
where $\ell$ is sampled according to Eq. (13), in which $n_e$
and $T_e$ should be understood as the electron density and temperature
in the present zone. With the existence of boundaries, 
$(x_1,y_1,z_1)$ could be in the neighboring zone or outside 
of the medium. In this case, one has to calculate the distances 
$s_i$ from  $(x_0,y_0,z_0)$ to various boundaries 
$B_i, (i=1,...,N)$, where $N$ is the number of boundaries surrounding
the zone under consideration. $s_i$ can be obtained by solving the 
equations describing the $i$th boundary
$$B_i(x,y,z)=0, \qquad i=1,...,N \eqno(22)$$

\noindent
where
$$\cases {x=x_0+s_i \omega_1 \cr 
          y=y_0+s_i \omega_2 \cr
          z=z_0+s_i \omega_3. } \eqno(23) $$

If $\ell$ is smaller than any of $s_1,...s_N$ so obtained, the photon
will remain in the same zone and scatter at the location $(x_1,y_1,z_1)$
on electrons at local temperature $T_e$.
But if $s_j$ is the minimum among $\ell$ and $s_1,...s_N$, the photon
will hit the boundary $B_j$. In this case one has to replace the photon
on the boundary at $(x,y,z)$ determined by Eq. (23) with $i=j$. 
With the new position on the boundary as $(x_0,y_0,z_0)$, one can begin 
another round of free path sampling with $n_e$ and $T_e$ of the zone 
the photon is entering but keeping the photon energy and direction
unchanged.
 
In the study of Gamma-ray spectra of Cygnus X-1 (Ling et al. 1997),
we developed a model where photons scatter in a two-layered spherical
plasma consisting of a high-temperature core and a cooler 
corona. The model was first proposed by Skibo and Dermer (1995) to 
interpret the X-ray spectral hardening at high energies observed in AGNs.
The boundary of the inner core is a sphere of radius $R_i$
while the boundaries of the outer shell are two spheres with radii
$R_i$ and $R_o$ respectively. For a photon in the core, the equation for    
the distance $s_1$ to its boundary is
$$s_1^2 + 2({\bf r_0} \cdot \hat{\bf \omega})s_1 - (R_i^2 - r_0^2) = 0,
\eqno (24) $$

\noindent
where ${\bf r_0} = (x_0,y_0,z_0)$ is the position vector of the photon;
$\hat{\bf \omega} = (\omega_1, \omega_2, \omega_3)$.
Similarly, the equations for a photon in the
outer shell are
$$\cases {s_1^2 + 2({\bf r_0} \cdot \hat{\bf \omega})s_1 - 
          (R_i^2 - r_0^2) = 0 \cr
s_2^2 + 2({\bf r_0} \cdot \hat{\bf \omega})s_2 - (R_o^2 - r_0^2) = 0 }.
\eqno (25) $$
\noindent 
Thus we have the following algorithm:
\begin{tabbing}
\= If \= $r_0 < R_i$, \\
\> \> Let $\delta = ({\bf r_0} \cdot \hat{\bf \omega})^2 + (R_i^2-r_0^2)$
and $s_1 = \sqrt{\delta} - ({\bf r_0} \cdot \hat{\bf \omega}).$ \\
\> \> If $\ell<s_1$, scatter at ${\bf r_1}={\bf r_0}+\ell\hat{\bf \omega}$. \\
\> \> Else reach boundary at ${\bf r_1}={\bf r_0}+s_1\hat{\bf \omega}$. \\
\> Else if $r_0 < R_o$, \\
\>    \> Let $\delta = ({\bf r_0} \cdot \hat{\bf \omega})^2 + 
(R_i^2-r_0^2)$. \\
\>\> If \= $\delta \geq 0$ and $ ({\bf r_0} \cdot \hat{\bf \omega}) < 0$, \\
\>\>\> Let $s_1 =-\sqrt{\delta}-({\bf r_0}\cdot \hat{\bf \omega})$. \\
\>\>\> If $\ell<s_1$, scatter  at ${\bf r_1}={\bf r_0}+
\ell\hat{\bf \omega}$. \\
\>\>\> Else reach boundary at ${\bf r_1}={\bf r_0}+s_1\hat{\bf \omega}$. \\
\>\> Else \\
\>\>\> Let $\delta = ({\bf r_0} \cdot \hat{\bf \omega})^2 + (R_o^2-r_0^2)$
and $s_2 = \sqrt{\delta} - ({\bf r_0} \cdot \hat{\bf \omega}).$ \\
\>\>\> If $\ell<s_2$, scatter  at ${\bf r_1}={\bf r_0}+
\ell\hat{\bf \omega}$. \\
\>\>\> Else escape.
\end{tabbing}

\noindent
Whenever the photon crosses the inner boundary, the plasma density and
temperature should be switched while the photon energy and direction kept
unchanged. 

In figure 3, we present the result of such a calculation
(solid curve) together with the observational data (Ling et al. 1997)
it was intended to fit. The data was from the blackhole candidate
Cygnus X-1 observed by the detector BATSE on board satellite 
Compton Gamma-Ray Observatory.
The fitting spectrum was obtained from a calculation with the two-layer
model described above, where temperature is $kT_e = 230$ keV for 
the inner core and 50 keV for the outer shell. The 
two zones are assumed to have the same electron
density and the inner core has a radius 0.36 in units of Thomson mean
free path, while the outer-shell radius is 1.3. The initial photons
have a blackbody temperature of 0.5 keV and injected into the medium
from outside. For comparison the best fit one can achieve by a 
single-zone plasma model is also presented (dashed curve). The 
model consists of a plasma sphere of radius 1.35 at $kT_e = 85$ keV. 
The reduced $\chi^2$ value is 2.6 for the single-temperature model as
compared to 1.0 for the double-layer model. It is seen that by
adding a hot central core to the Comptonization medium, the fit to
the high-energy part of the observed spectrum is significantly improved. 

\begin{figure}[ht]
\plotone{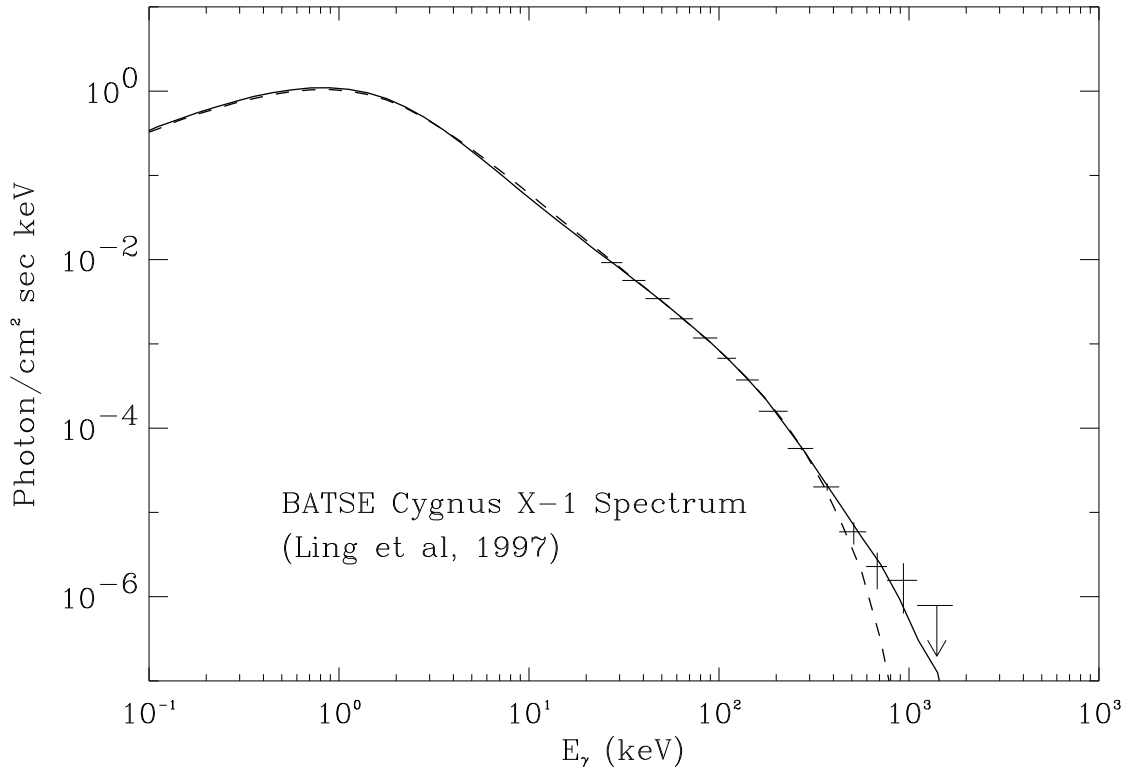}
\caption{
The energy spectra resulting from the double-layer
Comptonization media (solid curve) and singe-temperature sphere (dashed
curve. Both spectra are intended to fit the observational data from
the blackhole candidate Cygnus X-1 (Ling el al, 1997). }
\end{figure}

\section{Comptonization in Isothermal Media with Non-Uniform Density}

The media we considered so far are uniform, at least regionally, in density.
It was found necessary to investigate the Comptonization in the media with
non-uniform density profiles (Kazanas et al, 1997). In this section, we
present the treatment of two spherically symmetrical configuration
commonly found in astrophysical environment, one with electron density 
varying as $\rho^{-1}$ and the other as $\rho^{-3/2}$, where 
$\rho$ is the distance from the sphere center. The latter case represents
the density profile of a gas free-falling onto a central accreting
object under gravitational force (e.g. Narayan \& Yi 1994), 
while the former represents that of
an accreting gas with viscosity due to the interaction between the
gas and the outgoing photons (Kazanas et al, 1997). 

With density $n_e$ varying along the photon's path length $s$, 
the integration in Eq. (12) should be written as 
$$ I = \int_0^{\ell} n_e(s) \sigma_a ds, \eqno(26)$$

\noindent
where the dependence of $n_e$ on $s$ is given by 
$$n_e(s) = \cases{\displaystyle{{n_0 \rho_0} \over 
\sqrt{r_0^2+s^2+2sr_0\nu}} & {\rm for $\rho^{-1}~~$ profile},\cr 
\displaystyle{{n_0 \rho_0^{3/2}} \over 
{(r_0^2+s^2+2sr_0\nu)^{3/4}}} \quad 
& {\rm for $\rho^{-3/2}~~$ profile}, } \eqno(27)$$ 

\noindent 
where $\rho_0$ is the radius of the sphere within which the density 
profiles break down; $n_0$ is the electron density at this radius; 
$\nu=({\bf r_0} \cdot \hat{\bf \omega})/r_0$; ${\bf r_0}$ is the 
photon's position vector originated from the sphere center and 
$\hat{\bf \omega}$ its travel direction.

Substitute $n_e(s)$ in Eq. (27) into Eq. (26) and we obtain the
integration for $\rho^{-1}$ profile
$$ I = n_0 \rho_0 \sigma_a \ln \left[\frac{\ell -r_0 \nu + 
       \sqrt{\ell^2 + r_0^2 + 2\ell r_0 \nu}}{r_0 (1-\nu)}\right]. \eqno(28)$$ 

\noindent
Eq. (12) then becomes $I = -\ln \xi$. Solving this
equation for $\ell$, we obtain
$$ \ell = r_0 \displaystyle {{(1+\nu)\eta^2 + 2\nu \eta - (1-\nu)} 
          \over {2 \eta}}, \eqno(29)$$

\noindent
where $\eta = \exp(- \ln \xi/n_0 \rho_0 \sigma_a)$. Once a uniform random
$\xi$ is selected on $(0,1)$, $\ell$ is determined by Eq. (29).

For $\rho^{-3/2}$ density profile, the counterpart of Eq. (28) is
$$ I = n_0 \rho_0 \sigma_a \sqrt{\displaystyle {{2 \rho_0} \over {r_0
\sin \vartheta}}}~ [F(\varphi_{\ell}, \frac{1}{\sqrt{2}}) -
F(\varphi_0, \frac{1}{\sqrt{2}})], \eqno(30)$$

\noindent
where $F(\varphi, k)$ is the Legendre elliptic
integral of the 1st kind; $\sin \vartheta = \sqrt{1 - \nu^2}$;
$\varphi_0$ and $\varphi_{\ell}$ are given by 
$$\cases {\cos \varphi_0 = (1+ u_0^2)^{-1/4} \cr
          \cos \varphi_{\ell} = (1+ u_{\ell}^2)^{-1/4},} \eqno(31)$$

\noindent
and
$$\cases {u_0 = \displaystyle{{\cos \vartheta }\over {\sin \vartheta}} \cr
          u_{\ell} = \displaystyle{{\ell + r_0 \cos \vartheta}\over
                     {r_0 \sin \vartheta}}.}  \eqno(32)$$

\noindent
Substituting the integration into Eq. (12), we obtain
$$ F(\varphi_{\ell}, \frac{1}{\sqrt{2}}) = F(\varphi_0, \frac{1}{\sqrt{2}}) 
   - \frac{\ln \xi}{n_0 \rho_0 \sigma_a} 
   \sqrt{\frac{r_0 \sin \vartheta}{2 \rho_0}}, \eqno(33)$$

\noindent
where the right-hand side is a function of known variables. Call it
$f(\xi, r_0, \vartheta)$. Solve Eq. (33) and we obtain
$$\cos \varphi_{\ell} = {\rm cn}(f,\frac{1}{\sqrt{2}}), \eqno(34)$$ 

\noindent
where ${\rm cn}(f,k)$ is the Jacobian elliptic function, which is the inverse 
of the elliptic integral $F(\varphi_{\ell}, k)$.
Computer routines for both elliptic integral and Jacobian elliptic 
function are available in many mathematical libraries 
(e.g. Press et al, 1992). Finally, $\ell$ can be obtained from 
Eqs. (31) and (32)
$$\ell =r_0 \sin \vartheta \sqrt{{\rm cn}^{-4}(f,\frac{1}{\sqrt{2}}) -1}
        - r_0 \cos \vartheta. \eqno(35) $$
\noindent 
Once $\ell$ is available, one can use the algorithms described in the
previous section to determine if the photon scatters, escapes or hits the
boundary. 

We used $\ell$ given in Eqs. (29) and (35) 
to study the Comptonization in a two-layer spherical model similar 
to that in the last section but with the outer layer
having a $\rho^{-1}$ or $\rho^{-3/2}$ density profile. Specifically,
we assume the density in the outer shell is given by Eq. (27) with
$\rho_0 = R_i$ and the density of the inner core is constant $n_+$. 
It is found that the energy spectrum 
of the X-rays emerging from such system is different from
a uniform sphere with the same optical depth (Kazanas et al, 1997).
More importantly, with the decreasing density profiles, the outer
layer, or the ``atmosphere" can extend to a distance much greater than
the size a uniform system with the same optical depth can do, giving
rise to the time variation properties on a much greater time scale. 

As an example, we show in Figure 4 two light curves, or the
time-dependent fluxes, for X-ray photons escaping from two such  
core-atmosphere systems, one with $\rho^{-1}$ and the other with
$\rho^{-3/2}$ density profile for the atmospheres. 
For both density profiles, the temperature is 50 keV in the
atmosphere as well as in the core; the total optical depth is 2 in 
terms of Thomson scattering and the radius of 
the inner cores is assumed to be $2 \times 10^{-4}$ light seconds. 
The core density $n_+$ is slightly different from each other: 
$1.6 \times 10^{17}$ and $1.68 \times 10^{17} {\rm cm}^{-3}$ for $\rho^{-1}$ 
and $\rho^{-3/2}$ profiles respectively. For the outer atmospheres,
$n_0$ in Eq. (27) are $0.4 \times 10^{17} {\rm cm}^{-3}$ 
and $1.68 \times 10^{17} {\rm cm}^{-3}$ respectively. 
As a result the radii of the systems are 1.01 and 2.63 light
seconds respectively. 

\begin{figure}[ht]
\plotone{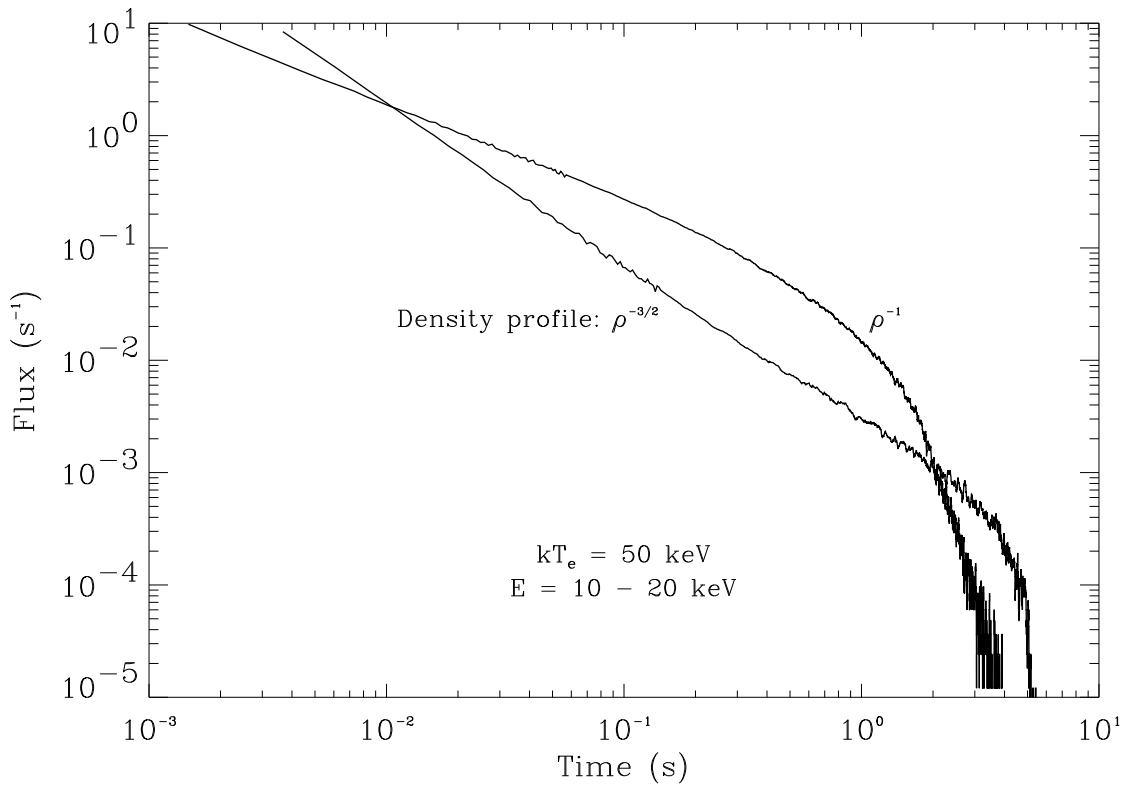}
\caption{
The light curves resulting from the core-atmosphere models.
The atmospheres have $\rho^{-1}$ and $\rho^{-3/2}$ density profiles
respectively.}
\end{figure}

Photons of a blackbody spectrum at temperature 2 keV are injected
at the center into the system. The Comptonized photons in the energy
range $ 10 - 20$ keV  are collected in terms of their escape time,
producing the light curves displayed in the figure.
It is seen that these light curves are power-laws extending to
the order of seconds followed by exponential cutoffs. The indices of the 
power-law are roughly 1 and 3/2 respectively, which was explained
in Kazanas et al (1997). This temporal behavior is greatly
different from the light curves from a uniform system, which decay
exponentially from the very beginning of the emissions (Hua \& 
Titarchuk, 1996). In addition, for a uniform system of the similar 
optical depth and an electron density of the order of $10^{16}$ or 
$10^{17} {\rm cm}^{-3}$, the characteristic decay time of the light 
curves will be $\sim 1$ millisecond. The implication of the 
prolonged power-law light curves resulting from the extended atmosphere models
for the interpretation of the recent X-ray observational data 
is discussed elsewhere (Kazanas et al. 1997, Hua et al. 1997). 
 
\section{Summary}  

We have shown that analytical method has intrinsic limitations in dealing
with Comptonization problem and Monte Carlo simulation provides a useful
alternative. We have 
introduced an efficient Monte Carlo method that can solve the
Comptonization problem in a truly relativistic way. The method was
further expanded to include the capabilities of dealing with 
Comptonization in the media where electron density and temperature
vary discontinuously from one region to the other and in the isothermal
media where density varies
continuously along photon paths. In addition to the examples
given above for its application, the method was also used in the
investigation of Compton scattering of gamma-ray photons in the
accretion disks near black hole candidates (Lingenfelter \& Hua,
1991) and in the Earth's atmosphere and the spacecraft material (Hua \& 
Lingenfelter, 1993). 

The author would like to thank R. E. Lingenfelter and R. Ramaty for
their long-term support and encouragement in the past decade during
which the technique described here was developed. The author also
wants to thank J. C. Ling, L. Titarchuk and D. Kazanas for valuable
discussions and NAS/NRC for support during the period of this
study.

\clearpage

\end{document}